\documentclass[prb,eqsecnum,twocolumn]{revtex4}
\usepackage{graphicx}%
\usepackage{dcolumn}
\usepackage{amsmath}

\begin{document}

\newcommand{\be}{\begin{equation}}
\newcommand{\ee}{\end{equation}}
\newcommand{\nin}{\noindent}

\newcommand{\femn}{\ensuremath{\text{Fe}_{50}\text{Mn}_{50}\,}}
\newcommand{\feni}{\ensuremath{\text{Ni}_{81}\text{Fe}_{19}\,}}
\newcommand{\spini}{\ensuremath{\mathbf{S}_{i}}}

\newcommand{\hres}{\ensuremath{H_{res}\,}}
\newcommand{\kone}{\ensuremath{K^{(1)}_{i}\,}}
\newcommand{\ktwo}{\ensuremath{K^{(2)}_{i}\,}}
\newcommand{\field}{\ensuremath{\mathbf{H}\,}}
\newcommand{\alm}{\ensuremath{\alpha_{M}\,}}
\newcommand{\Na}{\ensuremath{N_{a}\,}}
\newcommand{\Nb}{\ensuremath{N_{b}\,}}
\newcommand{\nn}{\ensuremath{N_{b}/N\,}}
\newcommand{\nf}{\ensuremath{N_{F}\,}}
\newcommand{\naf}{\ensuremath{N_{AF}\,}}
\newcommand{\efaf}{\ensuremath{\varepsilon_{F,AF}\,}}
\newcommand{\jaf}{\ensuremath{J_{AF}\,}}
\newcommand{\jfaf}{\ensuremath{J_{F,AF}\,}}
\newcommand{\hc}{\ensuremath{h_{c}\,}}
\newcommand{\hb}{\ensuremath{h_{b}\,}}
\newcommand{\ka}{\ensuremath{K_{a}\,}}
\newcommand{\kb}{\ensuremath{K_{b}\,}}
\newcommand{\keff}{\ensuremath{K^{eff}_{4}\,}}
\newcommand{\rtwo}{\ensuremath{R^{2}\,}}
\newcommand{\vh}{\ensuremath{\mathbf{h}\,}}
\newcommand{\vf}{\ensuremath{\mathbf{f}\,}}
\newcommand{\va}{\ensuremath{\mathbf{a}\,}}
\newcommand{\vb}{\ensuremath{\mathbf{b}\,}}
%\preprint{To be submitted to Physical Review B}

% Force line breaks with \\
\title[ ]{Induced four fold anisotropy and bias in compensated NiFe/FeMn double layers}

\author{T. Mewes}
\affiliation{%
Department of Physics, 1077 Smith Laboratory\\
Ohio State University\\
174 W 18th Ave, Columbus, OH 43210, USA\\
}

\author{B. Hillebrands}
\affiliation{%
Fachbereich Physik\\
Universit{\"a}t Kaiserslautern\\
Erwin Schr{\"o}dinger Str. 56, 67663 Kaiserslautern, Germany\\
}

\author{ R. L. Stamps}
\email{stamps@physics.uwa.edu.au}
\affiliation{%
School of Physics, M013\\
University of Western Australia\\
35 Stirling Highway, Crawley, WA 6009, Australia\\
}

\date{\today}

\begin{abstract}
A vector spin model is used to show how frustrations within a
multisublattice antiferromagnet such as FeMn can lead to four-fold
magnetic anisotropies acting on an exchange coupled ferromagnetic
film. Possibilities for the existence of exchange bias are
examined and shown to exist for the case of weak chemical disorder
at the interface in an otherwise perfect structure. A sensitive
dependence  on interlayer exchange is found for anisotropies
acting on the ferromagnet through the exchange coupling, and we
show that a wide range of anisotropies can appear even for a
perfect crystalline structure with an ideally flat interface.
\end{abstract}

%\pacs{75.70.-i}

\maketitle

%\tableofcontents
%\listoffigures
%\listoftables
\section{Introduction}

It is interesting to note that some of the technologically most
important exchange bias systems are also some of the most complex
and difficult to understand.  Antiferromagnetic metal compounds,
such as FeMn, can be used to pin ferromagnetic layers, making them
attractive for application in some devices. The unidirectional and
higher order magnetic anisotropies that appear in bilayers
containing FeMn seem to vary widely between experiments, and there
is no model yet capable of explaining the underlying microsopic
mechanisms.

One particularly intriguing unanswered question is how exchange
bias can occur in structures with compensated interfaces. Being
compensated, the ferromagnet spins in such structures interact
equally strongly with spins from all antiferromagnetic
sublattices. In the simplest approximation, there is no net
magnetic moment in the antiferromagnet for the ferromagnet to
couple to, and hence no way for the antiferromagnet to bias a
magnetization loop. A closer examination reveals that the
antiferromagnetic order at the interface is likely to be
frustrated, and a new configuration resulting in a small magnetic
moment at the interface should form \cite{koon:1997}. However it
has been shown that this so-called 'spin-flop' coupling is not of
its own accord able to support exchange bias during a
magnetization loop measurement \cite{schulthess:1998,butler:1999}.
%buttler:1999?

A number of considerations have been discussed that may explain
the existence of exchange bias in compensated interface
structures. It is highly likely that the interfaces are not
perfectly compensated due to defects in structure and chemical
composition. These imperfections of the interface give rise to
small numbers of uncompensated spins that can result in weak bias
shifts.

A second unanswered question is why the magnitude of the bias
shift is much smaller than the exchange field coupling the
ferromagnet and antiferromagnet. A number of possible explanations
have been put forward. These include the formation and pinning of
a partial domain wall near the interface\cite{mauri:1987}, the
pinning of domains and domain walls in the
antiferromagnet\cite{malozemoff:1988,milt:2000,nowak:1999}, and %nowak:1999?
interactions between grains in thin antiferromagnetic
films\cite{suess:2003}. In each explanation there are either
restrictions on film thickness or reliance on the existence of
structural or chemical disorder that make comparison to experiment
difficult.

As most models proposed so far focussed on explaining the
magnitude of the bias shift, they do not account for the
drastically increased coercivity observed in exchange bias
systems. While spin-flop coupling is not able to shift the
magnetization loop it does give rise to increased coercivity in
systems with a two sublattice antiferromagnet by inducing a
two-fold anisotropy in the ferromagnet \cite{butler:1999}.

In the present paper we demonstrate how frustrations within a
multisublattice antiferromagnet such as FeMn can lead to four-fold
magnetic anisotropies acting on the ferromagnet. As has been shown
previously the interplay between unidirectional and induced higher
order anisotropies acting on the ferromagnet not only causes an
increased coercivity but can also account for some of the complex
dependencies on applied field angle observed in exchange bias
systems
\cite{xi:1999,santos:1999,pokhil:2001,lai:2001,tang:2000,mewes:2002,pechan:2002}.

The possibility of exchange bias for a multisublattice
antiferromagnet is examined and shown for the case of weak
chemical disorder at the interface in an otherwise perfect
structure. Most importantly, we identify a sensitive dependence
for the anisotropies on interlayer exchange, and show that a wide
range of induced anisotropies can appear even for a perfect
crystalline structure with an ideally flat interface.

The paper is organized as follows. A vector spin model for FeMn is
introduced in the next section, followed by results obtained in
the limit of large interlayer exchange coupling. Interesting
possibilities for multiple configurations and effective
anisotropies appear with small interlayer coupling, and are
discussed in section III. The possibility of exchange bias in a
perfectly compensated system is discussed in section IV.

\section{Vector spin model and bulk equilibrium spin states }
Magnetic order in metallic antiferromagnets such as \femn is very
difficult to predict from first principles. One particularly
difficult aspect of the problem is the importance of contributions
that appear as magnetic anisotropies that are ultimately
determined by spin orbit coupling effects inside a crystalline
geometry. The problem of determining spin configurations
associated with exchange bias further requires consideration of a
large number of atoms in both the ferromagnet and antiferromagnet
films. These considerations make the problem very difficult to
approach from a quantum mechanical point of view.

In view of these difficulties, phenomenological approaches are
useful for developing insights into the problem. In this work we
represent the spin configuration using vector spins arranged at
lattice sites that represent the atomic ordering of a \feni/\femn
two film exchange coupled crystalline structure. Equilibrium spin
configurations are found as steady state solutions to coupled sets
of classical torque equations for each spin \spini\ located at
site $i$: \be \frac{d}{dt}\spini=\gamma \spini \times
\frac{\partial \cal H}{\partial\spini}+\alpha \spini \times \spini
\times \frac{\partial \cal H}{\partial\spini}. \label{eq:torque}
\ee

\nin The first term in this equation represents free precession of
spin \spini\ in its local field calculated from an appropriate
Hamiltonian, $\cal H$. The second term is dissipative with a form
chosen to preserve the length of the spin vector. The calculation
is intended to find equilibria only, and so the parameters
$\gamma$ and $\alpha$ are used only to control stability and
convergence so that the resulting dynamics does not represent
specific physical processes.

The Hamiltonian is chosen to be in the form of a Heisenberg spin
array with exchange interactions $J_{i,j}$, a Zeeman term for an
external applied field \field, and four-fold anisotropies \kone
and \ktwo. The exact form used is

\begin{eqnarray}
\nonumber \cal H =  &\sum_{i}& \big[ - g \mu_{B} \field \cdot \spini - 2 \sum_{\delta}J_{i,i+\delta}\spini\cdot\mathbf{S}_{i+\delta} \\
 &+&\kone \left(
S^{2}_{i,x}S^{2}_{i,y}+S^{2}_{i,x}S^{2}_{i,z}+S^{2}_{i,z}S^{2}_{i,y}
\right)\nonumber\\ &+&\ktwo S^{2}_{i,x}S^{2}_{i,y}S^{2}_{i,z}
\big]. \label{eq:hamil}
\end{eqnarray}

\nin The exchange sum is over nearest neighbors at sites
$i+\delta$, and the constants $g$ and $\mu_{B}$ are the Land\'e
factor and Bohr Magneton, respectively. Units are such that Eq.
\ref{eq:hamil} is an energy density.

\begin{figure}[!ht]

(a)\includegraphics[width=
0.8\linewidth,clip]{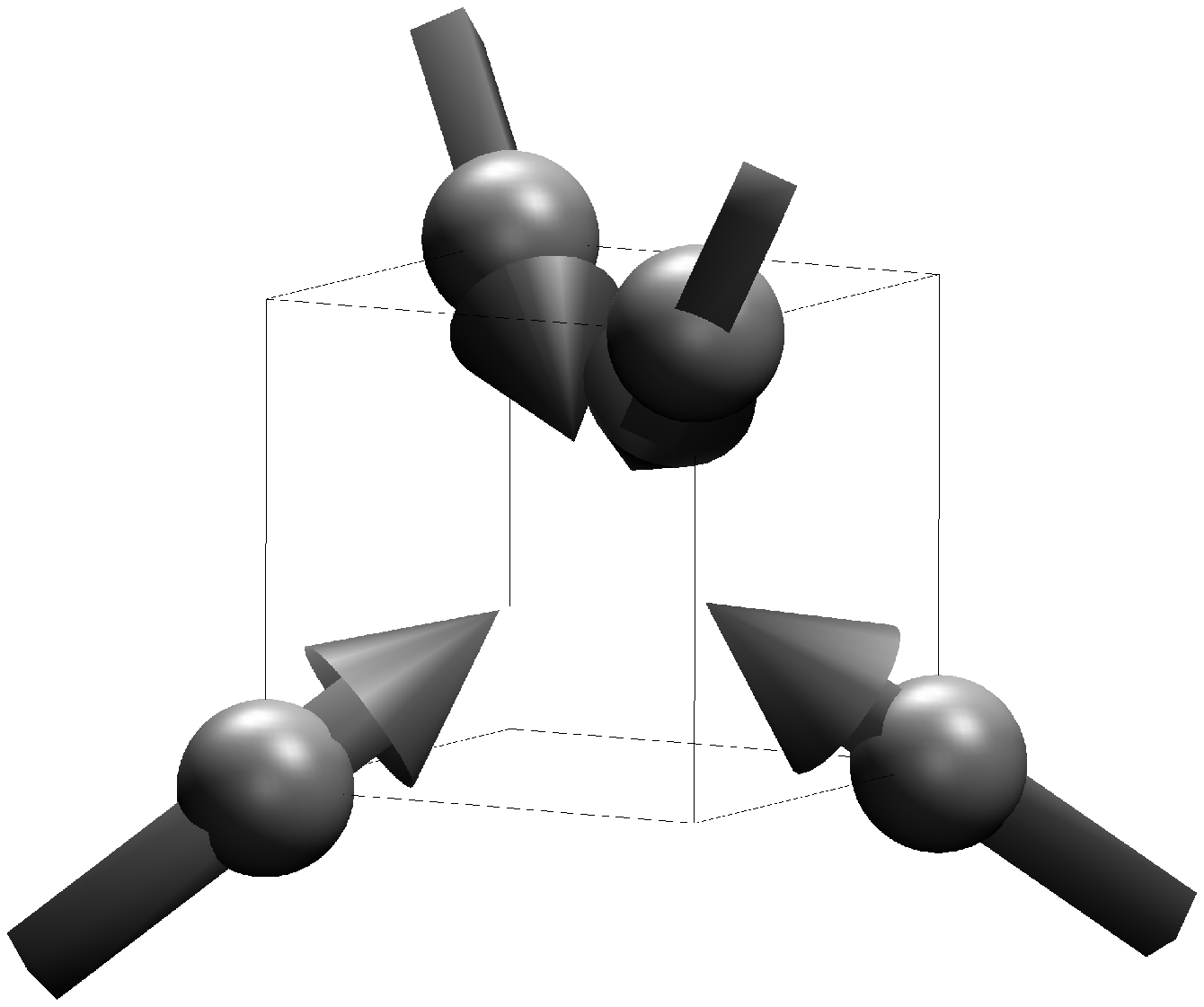}\hspace{0.5cm}
(b)\includegraphics[height=0.8\linewidth,clip]{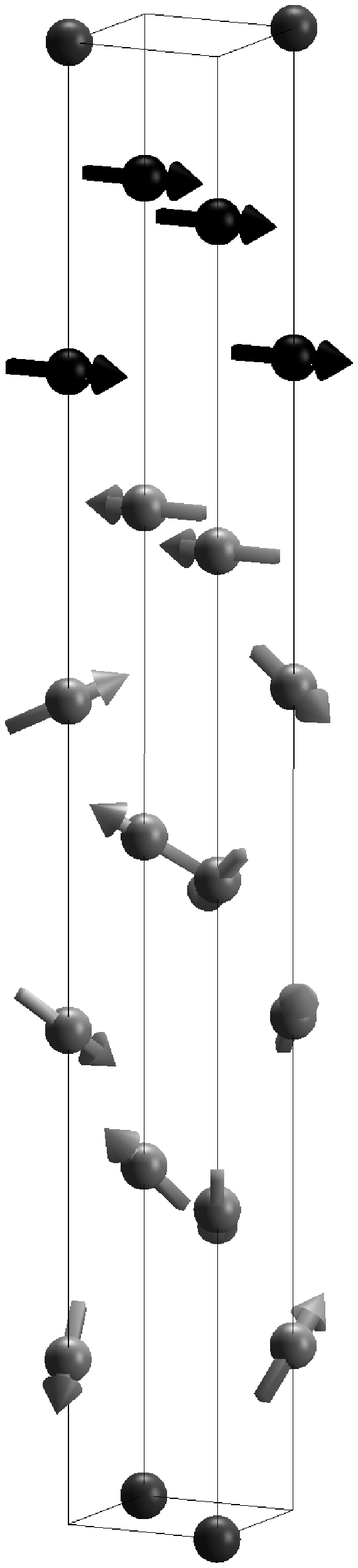} \caption{
The $3\cal Q$ phase for bulk \femn are shown in (a). An example
configuration of spins is shown in (b) for \femn exchange coupled
to two layers of \feni. The dark spheres indicate locations within
\feni. Periodic boundary conditions are assumed in all directions
in (a), and in directions parallel to the film planes in (b).}
\label{fig:TM3.1}
\end{figure}

All calculations reported here assume fcc lattice structure for
both ferromagnet and antiferromagnet. The anisotropy of the
antiferromagnet is choosen to mimic the $3\cal Q$ phase of an fcc
antiferromagnet. It is useful to note that there are three simple
stable equilibrium configurations, and it is not entirely clear
from experiment which is favored in
$Fe_{50}Mn_{50}$\cite{kouvel63,umebayashi66,kennedy87,schulthess99}.
Recent calculations suggest the $3\cal Q$ phase to be the stable
low energy configuration of bulk \femn
\cite{maclaren89a,maclaren89b}. The $3\cal Q$ phase is realised in
the vector spin model if the anisotropies fulfill both of the
conditions \kone/\ktwo$<-1/9$ and \kone/\ktwo$<-4/9$. This phase
has spins aligned along cube diagonals as illustrated in Fig.
\ref{fig:TM3.1} (a). All calculations were made using periodic
boundary conditions in all directions.

The spin structure calculated using the above model for \femn is
strongly affected by surfaces in a thin film geometry, due to the
broken translational symmetry at the interfaces This especially is
true when the antiferromagnetic film is exchange coupled to a
ferromagnet. An example is shown in Fig. \ref{fig:TM3.1} (b) where
two atomic layers of ferromagnetically coupled spins are exchange
coupled to an \femn film. Antiferromagnetic anisotropies are
\kone=$-|\jaf|/10$, and \ktwo=$|\jaf|/10$ where \jaf is the
exchange coupling in the antiferromagnet. No anisotropy is assumed
in the ferromagnet in order to mimic properties of \feni. The
ratio of ferromagnet to antiferromagnet exchange energies is set
at $J_{F}/|\jaf |=1.55$. These values represent exchange energies
consistent with the ratio of ordering temperatures of \feni and
\femn.

The calculation of a spin configuration is performed in analogy to
a field cooling procedure. The ferromagnet spins are aligned
parallel in plane at an angle \alm  made with respect to the
$(001)$ direction. The antiferromagnet spins are initially
oriented randomly. The equations of motion are integrated
numerically until the condition

\be \sum_{i}\frac{1}{|\spini|}\bigg| \frac{d\spini}{dt}\Delta
t\bigg|<\epsilon \label{eq:end} \ee

\nin is satisfied where $\epsilon$ is taken to be on the order of
$10^{-9}$.

An interesting feature of the spin configuration shown in Fig.
\ref{fig:TM3.1}(b) is the near complete antiparallel alignment of
spins directly at the interface due to the strong coupling between
the ferromagnet and the antiferromagnet assumed in this case. This
leads to a strong modification of the entire spin structure of the
antiferromagnet to the extend that it no longer resembles the
$3\cal Q$ structure. The ordering at and near the interface is
sensitive to the magnitude of the interlayer exchange coupling
\jfaf, and a number of different configurations with similar
energies are possible. It will be shown below that this has
consequences on the effective anisotropies acting on the
ferromagnet, and also on possible mechanisms for exchange bias.

\section{Induced anisotropy and coercivity}

The magnetic anisotropies in the antiferromagnet are communicated
to the ferromagnet through the interaction at their common
interface. Cases of strong exchange coupling across the interface
and weak exchange coupling show strikingly different features
because of how spins order in the antiferromagnet.

\subsection{Strong interlayer coupling}
The strong interlayer exchange responsible for the antiparallel
ordering of spins near the interface in Fig. \ref{fig:TM3.1}(b)
provides a mechanism for anisotropies to be induced in the
ferromagnet. If the orientation of the ferromagnet is changed
through application of an external applied field, spins in the
antiferromagnet will be rotated through anisotropy easy and hard
axes. This affects the total energy of the system, and in
particular can be represented by effective fields acting on the
ferromagnet. The energy per spin is defined for \nf ferromagnet
spins and \naf antiferromagnet spins in a unit cell of the two
film structure as

\be \efaf=\frac{E_{F,AF}}{\nf+\naf}, \label{eq:e_spin} \ee

\begin{figure}[!ht]
\includegraphics[width=0.8\linewidth,clip]{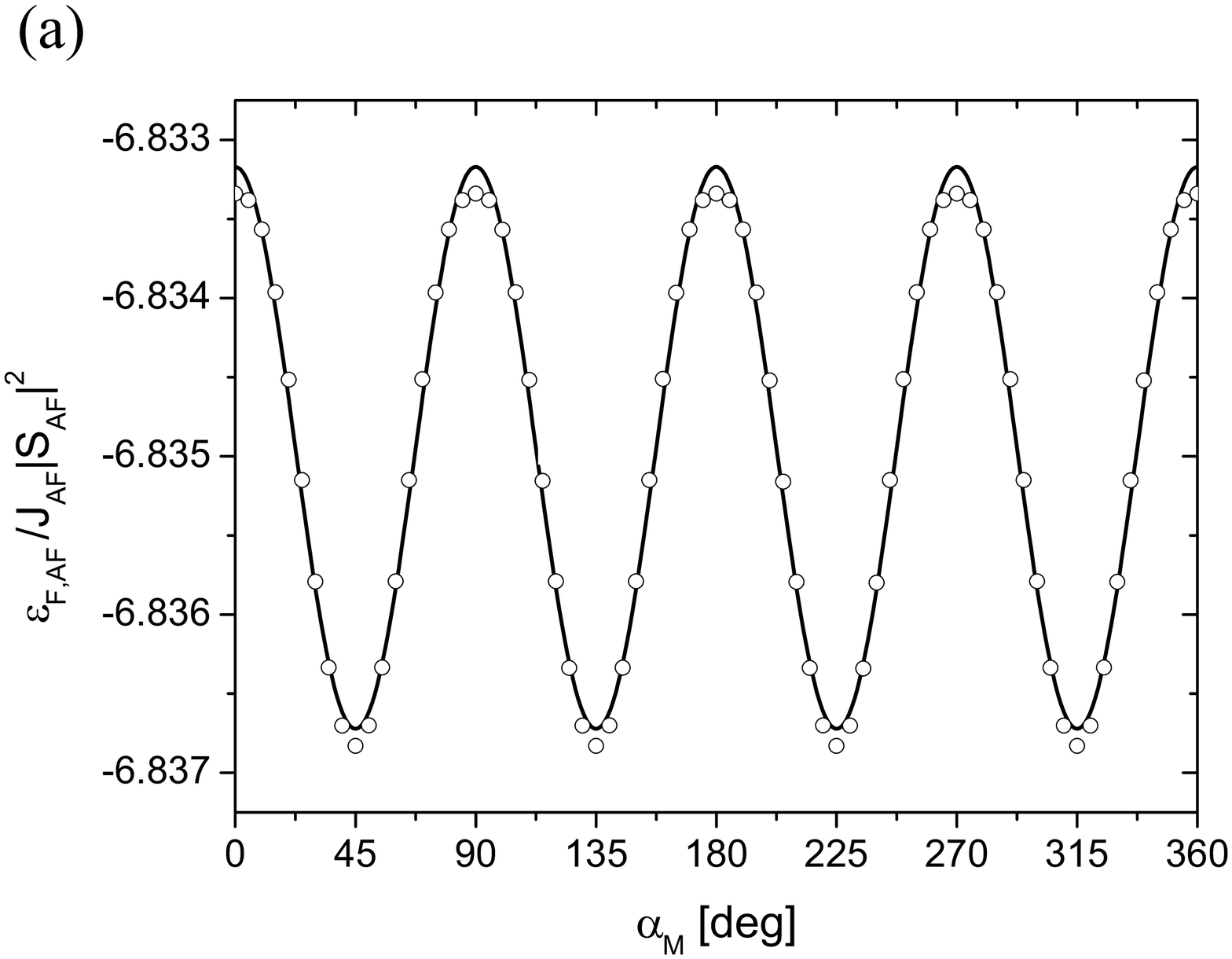}
\includegraphics[width=0.8\linewidth,clip]{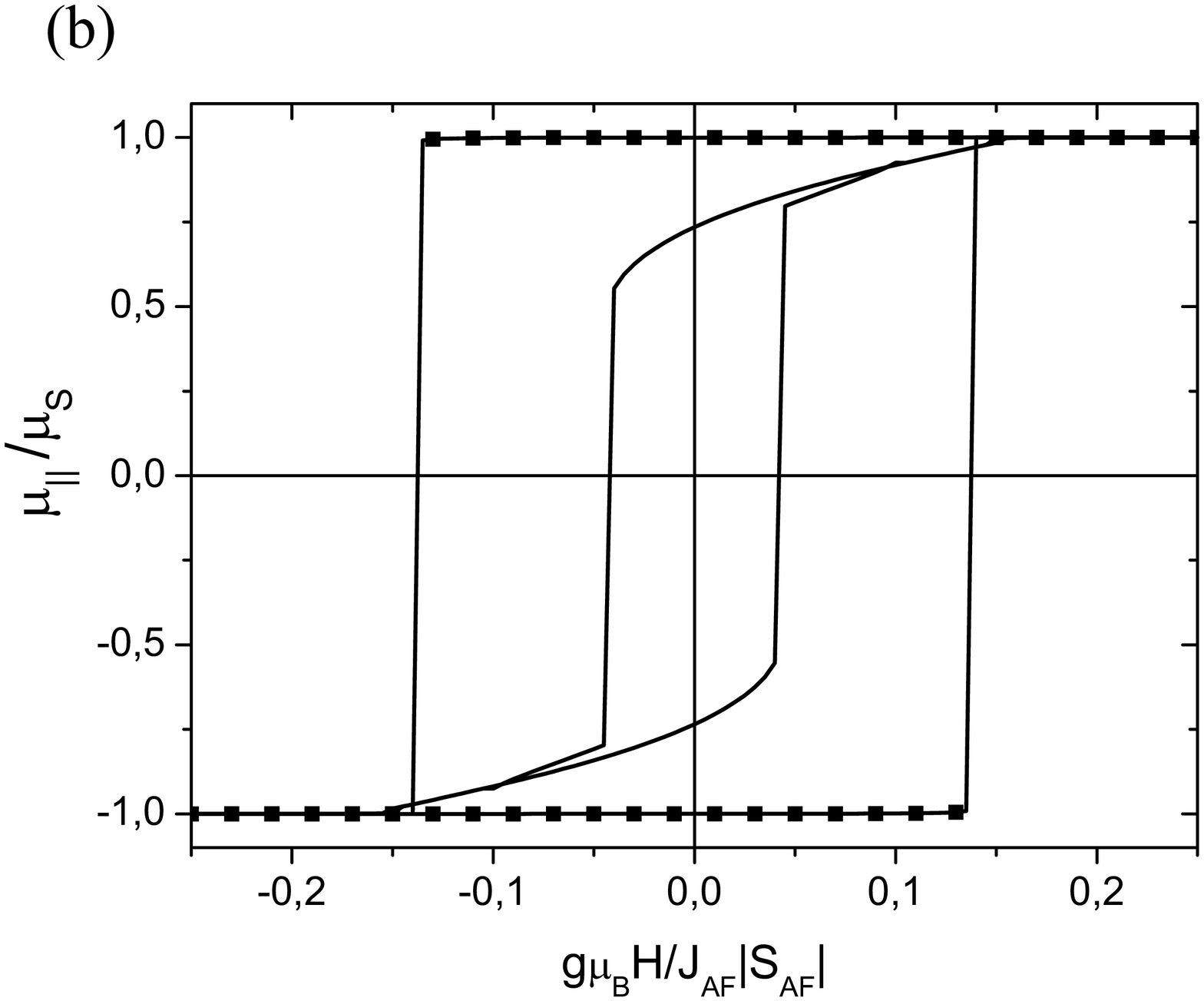} \caption{ Total energy per
spin is shown in (a) as a function of ferromagnet orientation. The
angle \alm is measured with respect to the $[100]$ axis. The
effective anisotropy displayed by the ferromagnet is four fold,
without a bias shift. An example magnetization loop is shown in
(b) for the applied field along the $[110]$ (filled circles) and
$[100]$ (solid line) directions.} \label{fig:TM3.4}
\end{figure}

\nin where $E_{F,AF}$ is the total energy of the two film system.
The energy density \efaf for an $\nf=4$ and $\naf=12$ system, as
depicted in Fig. \ref{fig:TM3.1}, is shown in Fig.
\ref{fig:TM3.4}(a). The energy is shown as a function of angle
\alm and is calculated by fixing the orientation of the
ferromagnet spins and allowing the antiferromagnet spins to relax
using the field cooling procedure described earlier. The field
cooling process was repeated for each orientation value of \alm.
The results of this procedure were compared to results from an
alternate method by which the antiferromagnet was field cooled
along \alm$=0$ and the energy calculated for each new \alm without
additional field cooling. The two different calculation procedures
produced identical results.

It is clear from Fig. \ref{fig:TM3.4}(a) that interlayer coupling
between the ferromagnet and antiferromagnet results in a total
energy for the system with four-fold symmetry for the in-plane
orientation of the ferromagnet.The solid line in Fig.
\ref{fig:TM3.4} is a fit using

\be \efaf=\keff \sin^2(\alm-\alpha_{4})\cos^2(\alm-\alpha_{4})
\label{eq:fourfold} \ee

\nin where $\alpha_{4}$ describes the orientation of the
anisotropy easy and hard axes with respect to the $[100]$
crystallographic direction.

Magnetization loops calculated for this set of \kone, \ktwo, and
\jfaf parameters show properties consistent with a simple four
fold anisotropy of the form in Eq. \ref{eq:fourfold}. An example
is shown in Fig. \ref{fig:TM3.4}(b) for a field applied along the
$[110]$ and $[100]$ directions. The magnetization is given in
units of Bohr Magneton and is defined as
$\mu_{S}=-g\mu_{B}\sum_{i}^{\nf}{|\spini|}$. The easy direction is
for $\alpha_{4}=\pi/4$. There is coercivity for the field along
the easy direction as would be expected due to the existence of
stable and metastable states for the ferromagnet parallel and
antiparallel to the applied field. Coercivity along the hard
direction exists because of the zero field remanent magnetization
aligned along an easy direction.

\subsection{Moderate and weak interlayer coupling}

\begin{figure}[!ht]
\includegraphics[width=0.8\linewidth]{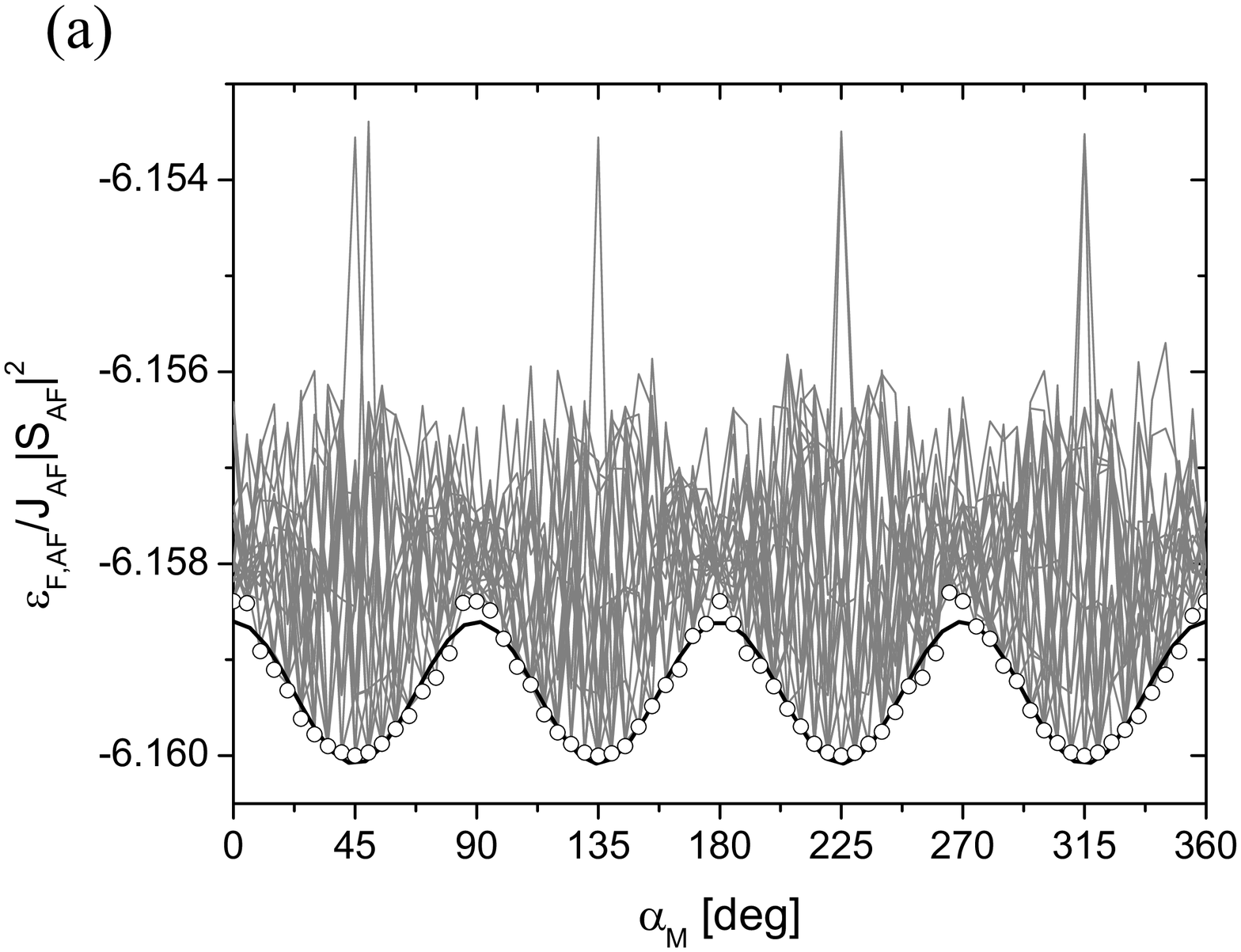}
\includegraphics[width=0.8\linewidth]{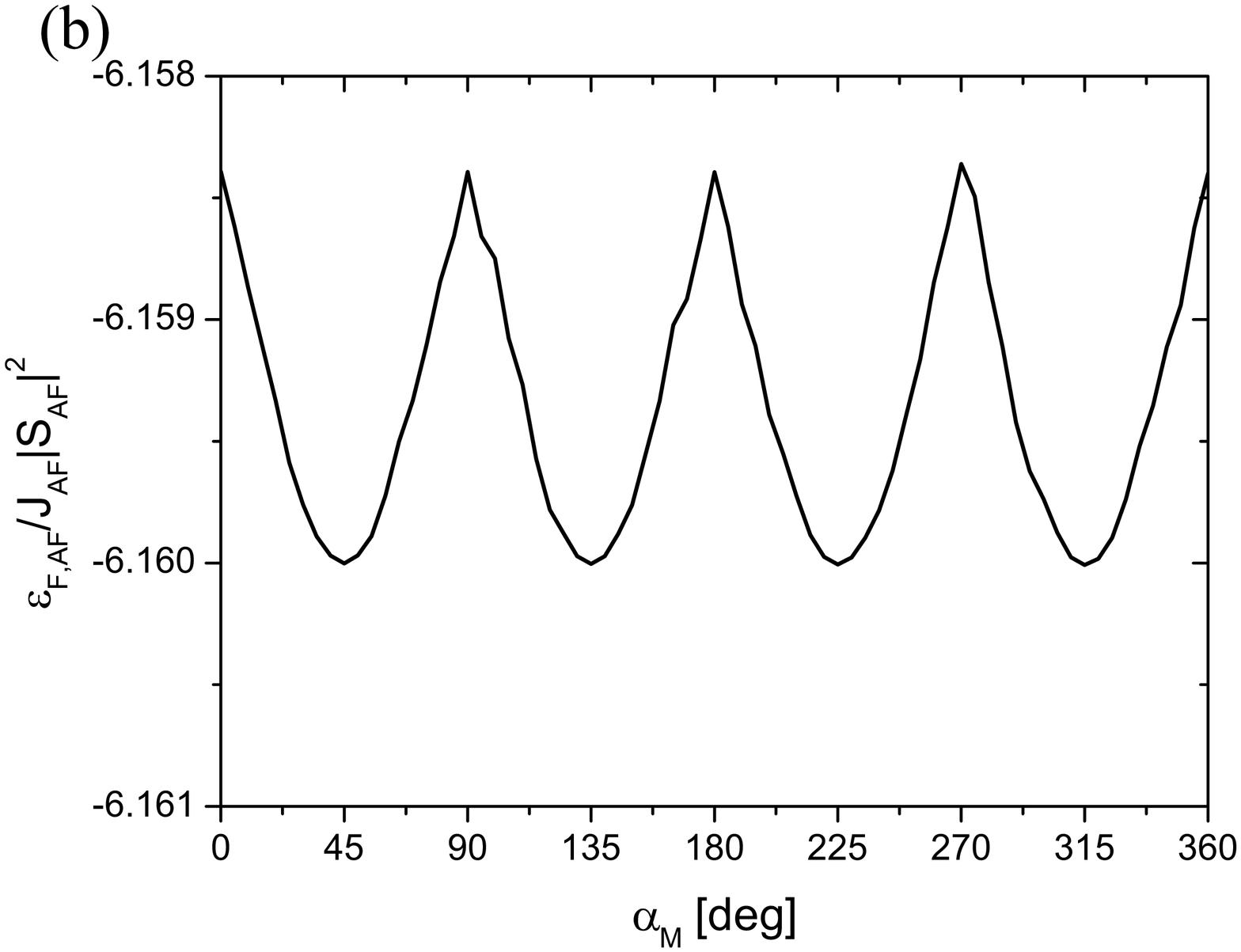}
\caption{ Total energy as a function of angle with \jfaf$=-0.5
|\jaf|$. Other parameters and structure are the same as used for
Fig. \ref{fig:TM3.4}. In (a), energies from thirty different
random initial configurations calculated at each angle are shown.
The spread in energies at each angle is due to the existence of
several metastable spin configurations within the antiferromagnet.
The lowest energies found are identified by open symbols in (a)
and plotted together in (b). } \label{fig:TM3.6}
\end{figure}

In the case of large \jfaf the magnetic behaviour of the
ferromagnet is dominated by the intra- and interlayer exchange
coupling and is well described by a simple four fold anisotropy.
When \jfaf is not large relative to \jaf, new possibilities for
metastable ordering within the antiferromagnet appear. Examination
of results from numerical solutions to Eq. \ref{eq:torque} reveal
multiple equilibrium configurations with comparable but different
energies. An example of this behavior is shown in Fig.
\ref{fig:TM3.6} where the the total energy \efaf is shown as a
function of angle for an interlayer exchange coupling
\jfaf$=-0.3|\jaf|$. The parameters and geometry are otherwise as
used for the example shown in Fig. \ref{fig:TM3.4}. These results
were generated using the field cooling procedure at each angle. At
each angle, thirty different random initial configurations for the
antiferromagnet spins were used, resulting in a spread of energies
as shown in Fig. \ref{fig:TM3.6}(a). The range of energies at each
angle represents a sampling of different possible spin
configurations.

The open symbols connected by the thick solid line in Fig.
\ref{fig:TM3.6}(a) are the lowest energies found during the
cooling process. In Fig. \ref{fig:TM3.6}(b) the lowest energies
are plotted separately to more clearly show the four-fold symmetry
of the ground state energies. Note that the energy \efaf displays
very sharp maxima and is only approximately consistent with the
\keff as given in Eq. \ref{eq:fourfold}.

The degree to which the lowest energy configurations result in an
anisotropy approximating that described by Eq. \ref{eq:fourfold}
is strongly dependent on the strength of the interlayer coupling
\jfaf. For some values of \jfaf, even the sign of the effective
anisotropy can change, representing a change in the orientation of
the easy and hard axes. The magnitude and sign of \keff saturates
to a fixed value for \jfaf larger than $1.25|\jaf|$. At these
large values, the antiferromagnet spins at the interface are
aligned collinear with respect to the ferromagnet spins, and
rotate rigidly with the ferromagnet as described in the previous
chapter.

\begin{figure}[!ht]
\includegraphics[width=0.8\linewidth,clip]{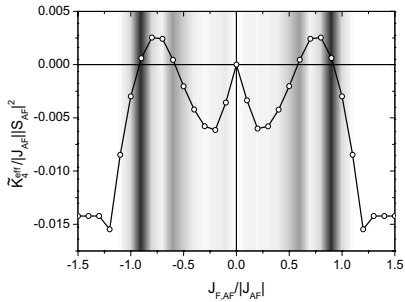}
\caption{ Effective anisotropy describing ground state energy
configurations as a function of  interlayer coupling \jfaf. The
degree to which the anisotropy is a simple four-fold type of the
form given in Eq. \ref{eq:fourfold} is indicated by grayscale
shading. White means that the description is very good, and black
means that the description is very poor. Note that the direction
of the easy and hard axes can change depending on the magnitude of
\jfaf. } \label{fig:TM3.7}
\end{figure}

A plot of the effective anisotropy determined by fitting the
ground state energy configurations to Eq. \ref{eq:fourfold} is
given in Fig. \ref{fig:TM3.7} as a function of interlayer coupling
\jfaf. The goodness of fit is quantified by the measure \rtwo
where $\rtwo=1$ is very good and denoted by lightly shaded
stripes, and very bad bits are denoted by dark gray stripes.

It is interesting to examine separately the ferromagnet and
antiferromagnet contributions to \efaf. In particular, the
ferromagnet component of the energy can have a different
dependence on orientation angle \alm than the total system energy.
A way of thinking of this is to consider the effective field
$\mathbf{h}_{eff}$ acting at a layer of spins within the structure
and calculate the associated energy
$-\spini\cdot\mathbf{h}_{eff}$. The field varies in magnitude and
direction throughout the structure, and can have a dependence on
\alm that also varies from layer to layer. A way of characterizing
this difference in a meaningful way is to identify extrema in the
energy of the ferromagnet calculated as a function of \alm, and
compare this to the extrema determined from the total energy
\efaf.

\begin{figure}[!ht]
\includegraphics[width=0.8\linewidth]{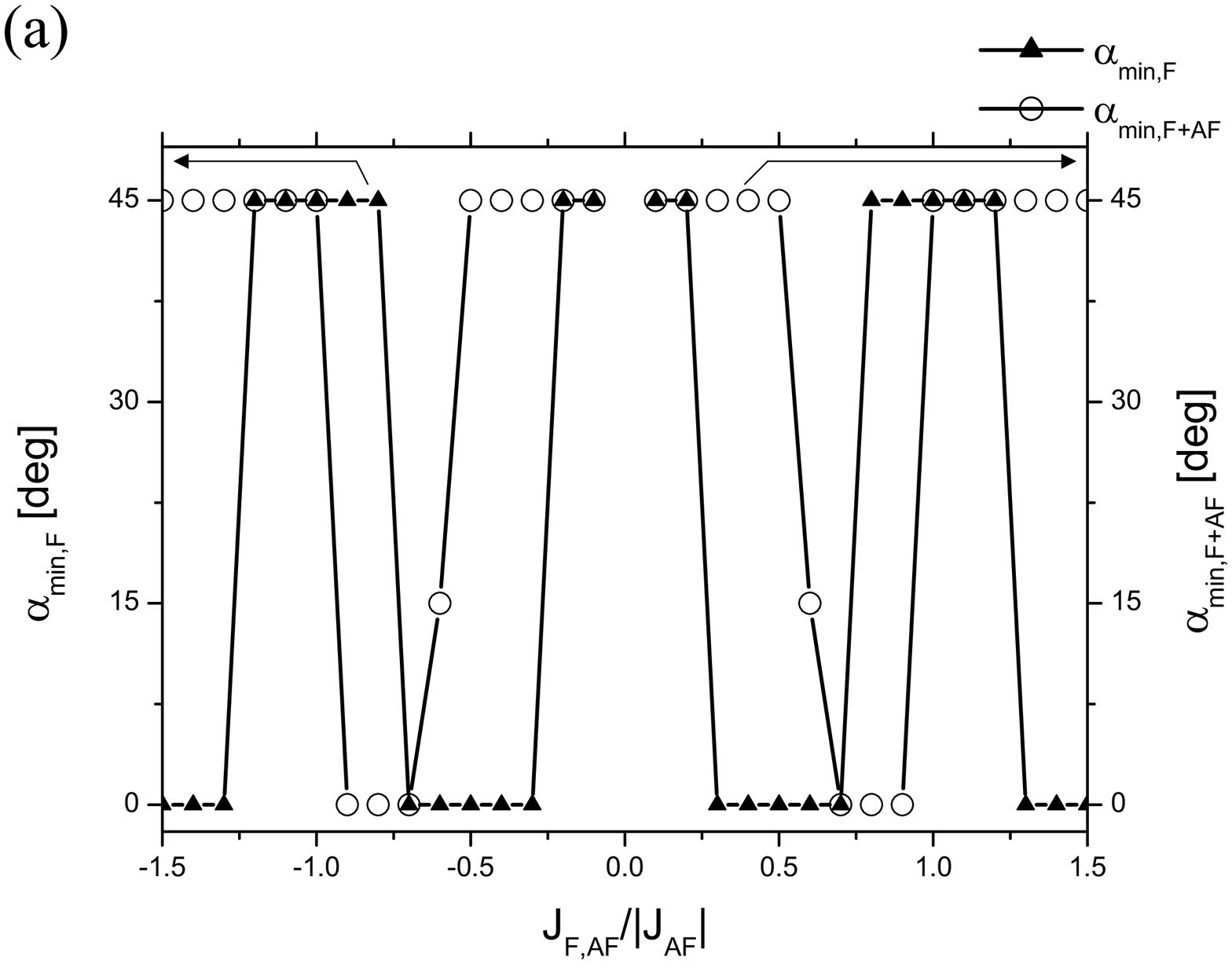}
\includegraphics[width=0.8\linewidth]{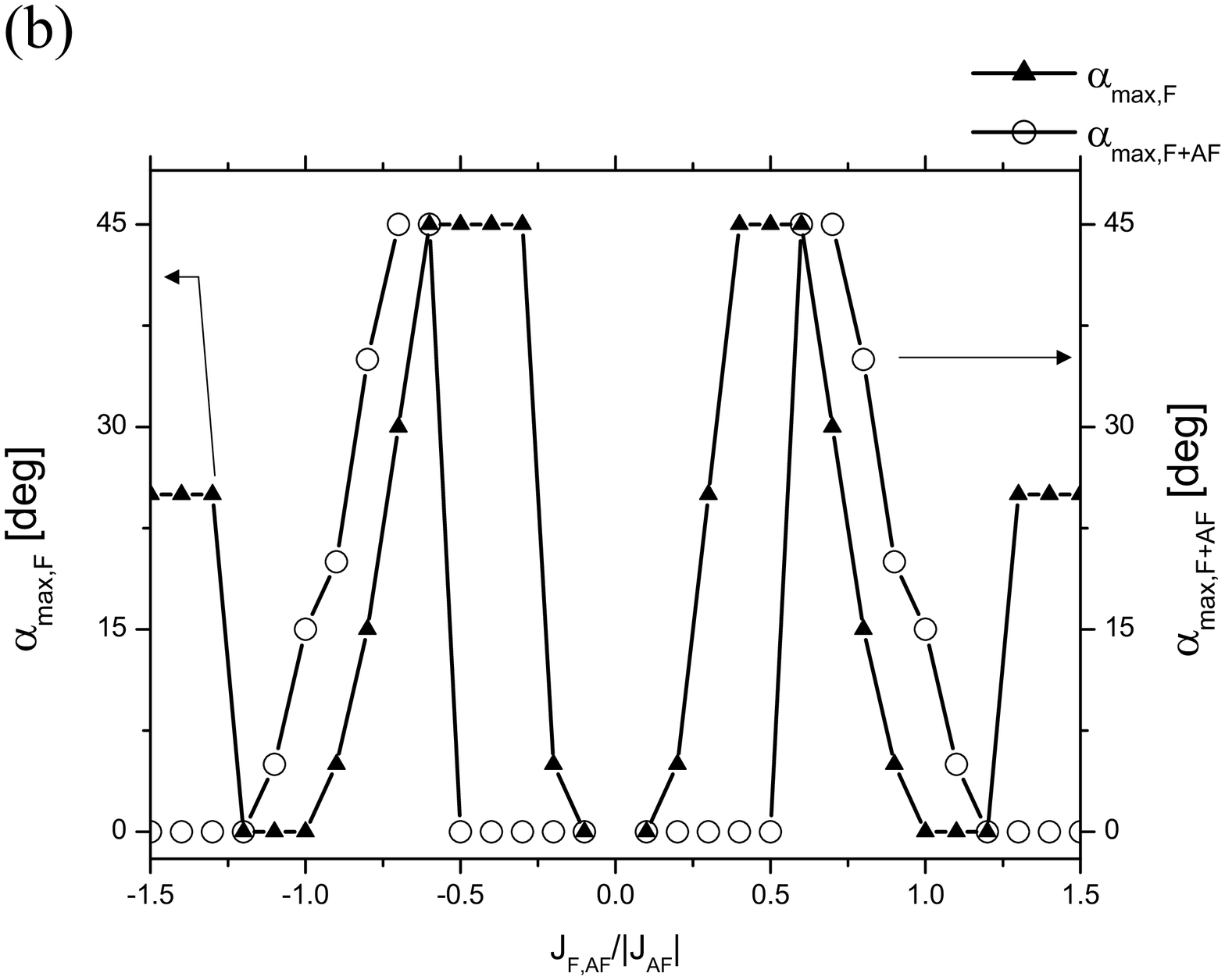}
\caption{Orientations for energy extrema as functions of \jfaf. In
(a), angles $\alpha_{min,F}$ represent orientations of the
ferromagnet spins where the energy of the ferromagnet spins has
minima, and the angles $\alpha_{min,F+AF}$ represent orientations
of the ferromagnet spins where the energy of the entire system has
minima. In (b) the angles $\alpha_{max,F}$ and $\alpha_{max,F+AF}$
are shown for the corresponding energy maxima.  }
\label{fig:TM3.1617}
\end{figure}

Results of this characterization are given in Fig.
\ref{fig:TM3.1617} as the angles \alm where minima (a) and maxima
(b) occur as functions of \jfaf. The angles $\alpha_{min,F}$
represent orientations of the ferromagnet spins where the energy
of the ferromagnet spins has minima, and the angles
$\alpha_{min,F+AF}$ represent orientations of the ferromagnet
spins where the energy of the entire system has minima. Likewise,
the angles $\alpha_{max,F}$ represent orientations of the
ferromagnet spins where the energy of the ferromagnet spins has
maxima, and the angles $\alpha_{max,F+AF}$ represent orientations
of the ferromagnet spins where the energy of the entire system has
maxima.

An interpretation of the minima is to assign different easy and
hard axes to the ferromagnet and system. In this view, the
alignment of the ferromagnet to an easy axis associated with the
system rather than the easy axis defined by the local effective
field acting on the ferromagnet is a clear indication of how order
in the antiferromagnet is involved in determining the magnetic
anisotropies observable through the ferromagnet. As an extreme
example, the large \jfaf limit shown in Fig. \ref{fig:TM3.1617}
has the ferromagnet 'easy' axis parallel to the system hard axis.
This occurs again at smaller values of \jfaf.

\begin{figure}[!ht]
\includegraphics[width=0.8\linewidth]{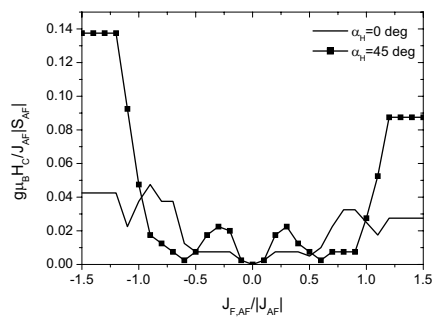}
\caption{Coercive fields for applied fields along the $[110]$ and
$[100]$ directions for different interlayer couplings \jfaf. The
closed symbols are for the applied field directed along the
$[110]$ and the open symbols are for the applied field directed
along $[100]$. } \label{fig:TM18}
\end{figure}

Coercivities calculated for magnetization loop simulations also
show curious behavior for some values of interlayer coupling.
Coercive fields for applied fields along the $[110]$ and $[100]$
directions are shown in Fig. \ref{fig:TM18}. The most curious
feature is the different coercivities produced with positive and
negative \jfaf. The reason for this is that the antiferromagnetic
film used for the calculation is very thin. Large values of
$|\jfaf|$ fully align the spins colinear with the ferromagnet and
contribute to the net magnetic moment. When $\jfaf>0$, the
contribution increases the Zeeman energy of the system in an
applied field, and when $\jfaf<0$, the contribution decreases the
Zeeman energy in an applied field. The easy axis coercivities are
therefore larger with negative \jfaf than with positive \jfaf.

\section{Mechanisms for exchange bias}

The anisotropies induced on the ferromagnet through interlayer
exchange coupling with the ferromagnet were not found to contain
any unidirectional contribution in any of the compensated
interface examples studied. This is consistent with previous
calculations demonstrating the inability of spin flop coupling at
compensated interfaces to support exchange bias shifts using
physically reasonable assumptions for anisotropy
fields\cite{stiles:2001,butler:1999}.%stiles 2001, butler:1999?

\subsection{Bias with interface defect}
It has been noted that small regions of uncompensated spins at the
interface can be sufficient to create exchange
bias\cite{nowak:1999,kouvel:1963,butler:1999,camley:1999a}. This possible %citation was \cite{nowak,kouvel,butler,camley}
mechanism for exchange bias is explored for the \feni/\femn model
discussed here. The unit cell of the structure with defect is
based on eight antiferromagnet atomic layers with eight spins in
each layer exchange coupled to a ferromagnet film consisting of
two atomic layers, also with eight spins per layer. Periodic
boundary conditions in the planes parallel to the interface as
used. The defect is represented by replacing the exchange
couplings and local anisotropies of one spin on the ferromagnet
side of the interface with values appropriate to the
antiferromagnet. In this way the unit cell of the structure has
$15$ ferromagnetically coupled spins, and $65$
antiferromagnetically coupled spins, with a small uncompensated
region at the interface.

\begin{figure}[!ht]
\includegraphics[width=0.8\linewidth]{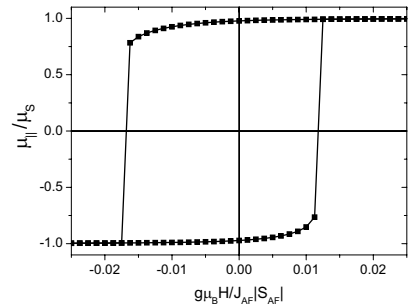}
\caption{Biased magnetization loop with one defect spin. The
parameters used are the same as those used for the unbiased loop
shown in Fig. \ref{fig:TM3.4}(b). The difference is that here the
unit cell has $\nf=15$, $\naf=65$, and one spin 'defect' located
at the interface providing a small net uncompensated
antiferromagnetic moment. \jfaf is $-0.5 |\jaf|$.}
\label{fig:TM3.8}
\end{figure}

Results from calculations for coupling between ferromagnet and
antiferromagnet spins with  \jfaf$=-0.5 |\jaf|$ are shown in Fig.
\ref{fig:TM3.8}.  A small negative bias of magnitude $0.005 |\jaf
S_{AF} |/ (g \mu_{B})$ appears. It is interesting to note that a
simple estimate of the shift would be to average the uncompensated
coupling energy $2$\jfaf over the number of ferromagnet spins,
giving a bias field of $0.07  |\jaf S_{AF} |/ (g \mu_{B})$. The
factor of fourteen discrepancy is because the spin order in the
antiferromagnet changes during magnetization in such a way as to
reduce the magnitude of the field necessary to align the
ferromagnet. This is analogous to the reduction in bias field
described by Mauri, et al\cite{mauri:1987} for exchange bias with
completely uncompensated interfaces.
\section{Summary}

A molecular field model of magnetic order in multi-sublattice
magnets, such as \femn, exchange coupled to a ferromagnet has been
examined. It has been shown that ordering of spins near the
surfaces and interfaces of the antiferromagnet are strongly
affected by exchange coupling, but that the intrinsic four-fold
anisotropy of the antiferromagnet is still induced into the
ferromagnet. Through exchange coupling, the ferromagnet spins
experience a four-fold anisotropy with a magnitude sensitive to
the strength of the interlayer coupling. An important point is
that the nature of the anisotropy is also dependent on the
strength of the coupling, and is only strictly of a simple four
fold form in the case of strong interlayer coupling.

Weak and moderate values of interlayer coupling lead to an
additional interesting effect on the induced anisotropy. Rather
than a single well defined order in the antiferromagnet, a number
of metastable configurations appear. The lowest energy
configurations lead to an induced anisotropy with four-fold
symmetry.  The degree to which the four-fold anisotropy associated
with the lowest energy configurations is described by a simple
sinusoid depends on the strength of the interlayer exchange
coupling and is only approximate at best for some values of the
exchange.

Exchange bias shifts for perfectly compensated interfaces assuming
the low energy spin configuration was not found. Instead, an
exchange bias could be created by allowing the interface to have
some small degree of uncompensation through introduction of a
defect in the interface structure. These findings are consistent
with previous work showing that a spin-flop configuration at a
compensated interface is incapable of producing a shifted
magnetization loop, whereas mixed interfaces with some amount of
uncompensation present can shift the magnetization loop.

\begin{acknowledgments}
TM and RLS acknowledge support by the Australian Research Council
(Discovery Grant and IREX).
\end{acknowledgments}
{}

%\bibliography{rstm_eb}
%\bibliography{len_eb}
%\bibliography{EBReview}

%\nocite{*}

\end{document}